\documentclass[useAMS,usenatbib]{mn2e}
\usepackage[percent]{overpic}
\usepackage{times}
\usepackage{graphicx}
\usepackage{amsmath}
\usepackage{mathabx}
\usepackage{subfigure}
\usepackage{natbib}
\usepackage{txfonts}
\usepackage{journals}
\usepackage{xcolor}
\usepackage{array}
\usepackage{makecell}

\newcommand{\HII}{$\rm H~{\scriptstyle II}$}
\newcommand{\HI}{$\rm H~{\scriptstyle I}$}
\newcommand{\Ha}{\ensuremath{\mathrm{H\alpha}}}

\newcommand{\kms}{$\rm km\,s^{-1}$}

\newcommand{\gb}{$G_{\rm BP}$}
\newcommand{\gr}{$G_{\rm RP}$}

\newcommand{\ebr}{$E(G_{\rm BP}-G_{\rm RP})$}

\title[Spiral structure from OB stars]
{The Galactic spiral structure as revealed by O- and early B-type stars} 

\author[B.Q. Chen et al.]
{B.-Q. Chen,$^{1}$\thanks{E-mail:
bchen@ynu.edu.cn (BQC); x.liu@ynu.edu.cn (XWL).}
 Y. Huang,$^{1}$
 L.-G. Hou,$^2$
 H. Tian,$^3$\thanks{LAMOST Fellow}
 G.-X. Li,$^1$
 H.-B. Yuan,$^4$
 H.-F. Wang,$^{1,5}$\footnotemark[2]
   \newauthor
 C. Wang,$^6$\footnotemark[2]
 Z.-J. Tian,$^{7}$
and  X.-W. Liu$^{1}$\footnotemark[1]
\\
$^{1}$South-Western Institute for Astronomy Research, Yunnan University, Kunming, Yunnan 650091, P.\,R.\,China\\
$^{2}$National Astronomical Observatories, Chinese Academy of Sciences, Beijing 100101, P.\,R.\,China \\
$^{3}$Key Laboratory of Optical Astronomy, National Astronomical Observatories, 
Chinese Academy of Sciences, Beijing 100012, P.\,R.\,China \\
$^{4}$Department of Astronomy, Beijing Normal University, Beijing 100875, P.\,R.\,China\\
$^{5}$Department of Astronomy, China West Normal University, Nanchong 637009,  P.\,R.\,China\\
$^{6}$Department of Astronomy, Peking University, Beijing 100871, P.\,R.\,China\\
$^{7}$Department of Astronomy, Yunnan University, Kunming, Yunnan 650091, P.\,R.\,China\\
 }

\begin{document}

\date{Accepted ???. Received ???; in original form ???}

\pagerange{\pageref{firstpage}--\pageref{lastpage}} \pubyear{2016}
\maketitle
\label{firstpage}

\begin{abstract}
We investigate the morphology and kinematics of the Galactic spiral structure based on a new 
sample of O- and early B-type stars. We select 6,858 highly confident OB star candidates from the 
combined data of the VST Photometric \Ha\ Survey Data Release 2 (VPHAS+ DR2) and the 
Gaia Data Release 2 (Gaia DR2).  Together with the O-B2 stars from the literature, 
we build a sample consisting of 14,880 O- and early B-type stars, all with Gaia
parallax uncertainties smaller than 20\,per\,cent. The new sample,
hitherto the largest one of O- and early B-type stars with robust 
distance and proper motion estimates, 
covers the Galactic plane of distances up to $\sim$ 6\,kpc from the Sun.
The sample allows us to examine the morphology of  
the Scutum, Sagittarius, Local and Perseus Arms in great detail. 
The spiral structure of the Milky Way as traced by O- and early B-type stars shows flocculent patterns.
Accurate structure parameters,
as well as the means and dispersions of the vertical velocity distributions of the individual spiral arms are presented. 
\end{abstract}

\begin{keywords}
stars: early-type -- Galaxy: disk -- Galaxy: structure  
\end{keywords}

\section{Introduction}

The Galactic spiral structure is an important feature 
when studying the characteristics and properties of the Milky Way disk. 
However, we still do not know its exact structure, due to 
the difficulties of obtaining accurate distances of the tracers \citep{Vallee2017, Xu2018b}.

Most of the current Galactic spiral models are based on the kinematic distances
of ionized \HII\ regions \citep{Georgelin1976, Downes1980, Caswell1987, Russeil2003, Hou2009},
\HI\ neutral gas \citep{Kerr1969, Weaver1970, Mcclure2004, Kalberla2009, Koo2017} and
CO molecular clouds \citep{Solomon1989, May1997, Heyer2001, Dame2011, Hou2014, Sun2015, Sun2017}. However, 
due to the large errors of those kinematic distances, the identification of the spiral arms
suffers from inevitable large ambiguity. Up to now, there is still no consensus on the 
number of arms, their locations and properties.

Precise distance is of paramount importance for tracing the structure of the
Galactic spiral arms. The accurate distance measurements of masers from the 
Very Long Baseline Interferometry (VLBI) have been vital to disentangle
the spiral arms, spurs and arm branches \citep{Xu2006, Honma2007, Reid2014}.
However, there are accurate trigonometric parallax observations 
for only about 100 masers, mostly in the first and second quadrants of the disk. 

The second Gaia data release (Gaia DR2; \citealt{Gaia2018}) has provided unprecedented high quality astrometric data
for over one billion stars. The uncertainties of Gaia parallaxes vary between 0.04 and  0.1\,mas for sources
of $G$ magnitudes between 8 and 18\,mag, yielding parallaxes accurate to 
20\,per\,cent out to distance of $\sim$ 5\,kpc \citep{Lindegren2018}. 
Based on the Gaia DR2 parallaxes, \citet{Xu2018} and \citet{Xu2018b} have studied the morphology of 
the Galactic spiral arm structure within 3\,kpc of the Sun using a sample of OB stars from \citet{Reed2003}.
They find that the nearby spiral structure unraveled by this sample agrees well with that revealed by 
the VLBI masers and extends to the fourth quadrant. They also find new spur-like structures between the major arms.

The Reed OB-star sample that \citet{Xu2018} and \citet{Xu2018b} employed is mainly
confined to objects brighter than 12th magnitude, and is complete
only to $\sim$ 2 kpc. To investigate the Galactic spiral pattern in an even larger spatial volume, one needs a new sample of 
OB-star candidates of a much deeper limiting magnitude. 
\citet{Skiff2014} have presented a catalogue of stars with spectral classifications collected from the literature.
Some OB stars in the catalogue are fainter than 20th magnitude. 
The Galactic O-Star Catalog (GOSC; \citealt{Maz2013, Maz2016}) 
contains several hundreds O and other early-type (B/A) stars
of magnitudes up to $B$ $\sim$ 16\,mag, selected from the Galactic O-star spectroscopic survey (GOSSS; \citealt{Maz2011}).
Based on a pure photometric selection algorithm (initiated by \citealt{Johnson1953}), 
\citet{Mohr2015} and \citet{Mohr2017} have selected over 5,000 O- and early B-type star candidates 
down to a limiting magnitude of $\sim$ 20th magnitude using
the photometric data from the VST Photometric \Ha\ Survey (VPHAS+; \citealt{Drew2014}) for a sky area 
of 42\,deg$^2$. Their method is proved to be of high efficiency and purity. A follow-up spectroscopic campaign
of 276 candidates confirms that 97\,per\,cent of them are bona fide OB stars \citep{Mohr2017}.
Most recently, \citet{Liu2019} present a catalogue of about 16,000 OB stars identified in the LAMOST spectroscopic
surveys \citep{Deng2012, Liu2014, Yuan2015}. However, most of them are late B-type (B4 or later) stars and there are only about 
300 O-B2 stars.

The VPHAS+ data release 2 (VPHAS+ DR2; \citealt{Drew2014}) have presented  
$u,~g,~r,~i$ and \Ha\ photometry of over 300 million objects covering 629\,deg$^2$ regions of the Galactic plane. 
Together with the Gaia DR2, they provide us a great
opportunity to select a new sample of O- and early B-type stars and explore the Galactic spiral structure
up to $\sim$ 6\,kpc from the Sun. 
In this paper, we have selected a deep sample of OB-star candidates from the VPHAS+ DR2
and Gaia DR2 catalogues. Combined with the O-B2 stars available from the literature, we examine the Galactic spiral
structure in unprecedented detail.

The paper is structured as following. In Section 2, we present the
relevant VPHAS+ DR2 and Gaia DR2 data. Section 3 describes
the new OB-star sample.
In Section 4, we present our main results which are
discussed and summarized in Section 5. 

\section{VPHAS+ and Gaia}

The new OB star candidates are selected from the VPHAS+ DR2 and Gaia DR2 catalogues. 

The VPHAS+ Survey \citep{Drew2014} collected images in the SDSS $u,~g,~r,~i$ broad bands and the 
\Ha\ narrow band using the OmegaCAM imager \citep{Kuijken2011} on the VLT Survey Telescope (VST). 
The survey is designed to cover $\sim$ 2000\,deg$^2$ of the Galactic plane in the southern hemisphere of Galactic latitude 
$|b|$ $<$ 5\degr\ and Galactic longitude
$-$150\degr\ $<$ $l$ $<$ 40\degr, and a Galactic bulge extension of $|b|$ $<$ 10\degr\ near 
the Galactic centre. The VPHAS+ DR2 \citep{Drew2014}, released in 2016, contains 
PSF and aperture photometry of $\sim$ 319 million point-like sources covering 629\,deg$^2$ of the planned VPHAS+
footprint.  Typically, a signal-to-noise ratio S/N = 5 cut corresponds to a limiting magnitude of about 22\,mag in the 
$u,~g$ and $r$ bands. The photometric calibration of VPHAS+ DR2 is 
consistent with that of the SDSS within an accuracy of 0.05\,mag (r.m.s error) for $g$ and $r$ bands. 

To exclude contaminations of sub- and over-luminous OB stars and stars of spectral types B4 and later, we combine the VPHAS+ photometry with
the Gaia DR2 data \citep{Gaia2018}. 
Gaia DR2 provides high precision photometric measurements of over 1.4 billion sources in $G$, \gb\ and \gr\ bands.
The Gaia $G$ band covers the entire optical wavelength ranging between 330 and 1050\,nm. 
The \gb\ (330 - 680\,nm) and \gr\ (630 - 1050\,nm) magnitudes are derived from
the Gaia low resolution spectrophotometric measurements.
The internal validation shows that the Gaia DR2 calibration uncertainties for the $G$, \gb\ and \gr\ bands are 
2, 5 and 3\,mmag, respectively.
Gaia DR2 also releases high-quality parallax and proper motion measurements of 1.3 billion sources.
The parallax uncertainties are around 0.04\,mas for bright sources of $G$ $<$ 14\,mag
and around 0.1\,mas for sources of $G$ $\sim$ 18\,mag.

\section{The OB-star sample}

\subsection{OB star candidates from VPHAS+ and Gaia}

\begin{table*}
   \centering
  \caption{Number of stars selected in the individual steps of the OB-star selection procedure in the current work.}
  \begin{tabular}{cc}
  \hline
  \hline
Description & Numbers   \\
\hline
{The initial VPHAS+ DR2 Sample} & {13,201,826} \\
{Sample selected from the VPHAS+ colour-colour diagram} & {91,688} \\
{Sample with the Gaia photometric and astrometric cuts} & {26,936} \\
\hline
\makecell[r]{The VPHAS+ OB star and candidates} & {6,858} \\
\makecell[r]{White dwarf contaminator} & {786} \\
\makecell[r]{Sub-dwarf contaminator} & {1,386} \\
\makecell[r]{B4- and later-type star contaminator} & {17,906} \\
\hline
{O-B2 stars from the literature} & {7,441 (Skiff), 442 (Ma{\'{\i}}z Apell{\'a}niz et al.) and  931 (Huang et al.)} \\
{Common stars between the VPHAS+ sample and those from the literature} & 
                  {93 (Skiff), 2 (Ma{\'{\i}}z Apell{\'a}niz et al.) and  3 (Huang et al.)} \\
{The combined sample of OB stars and candidates} & {14,880} \\
\hline
\end{tabular} 
\end{table*}

\begin{figure}
    \centering
  \includegraphics[width=0.49\textwidth]{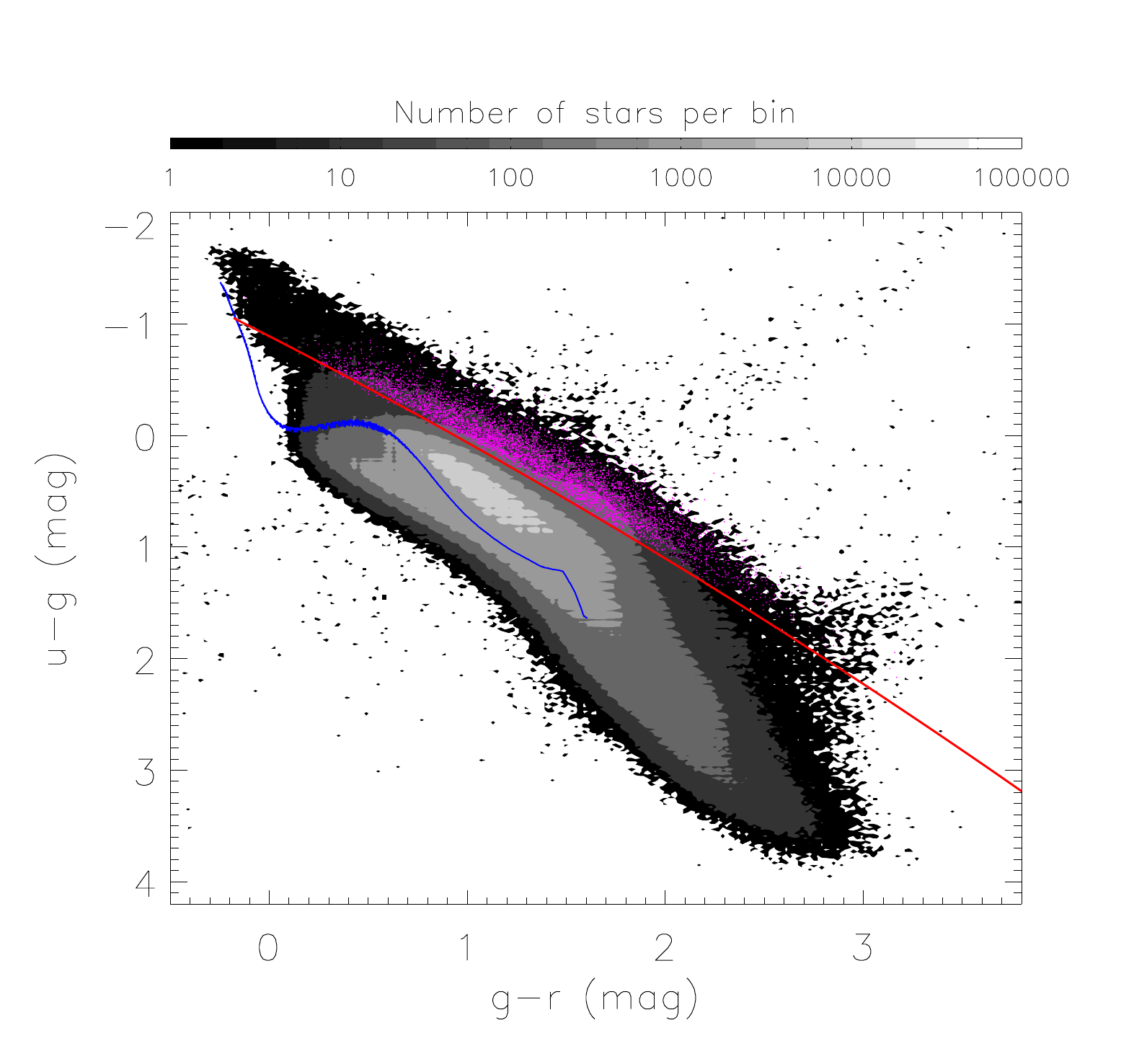}
  \caption{Distribution of stars selected from the VPHAS+ DR2 catalogue in the ($u-g$)  versus ($g-r$) 
  plane. The blue curve represents the stellar locus from the PARSEC isochrones \citep{Marigo2017}
  and the red line shows the reddening curve of the B3V stars from \citet{Drew2014}.
Pink dots are O-B2 candidates from \citet{Mohr2017}. }
  \label{ccd}
\end{figure}

 \begin{figure*}
    \centering
  \includegraphics[width=0.99\textwidth]{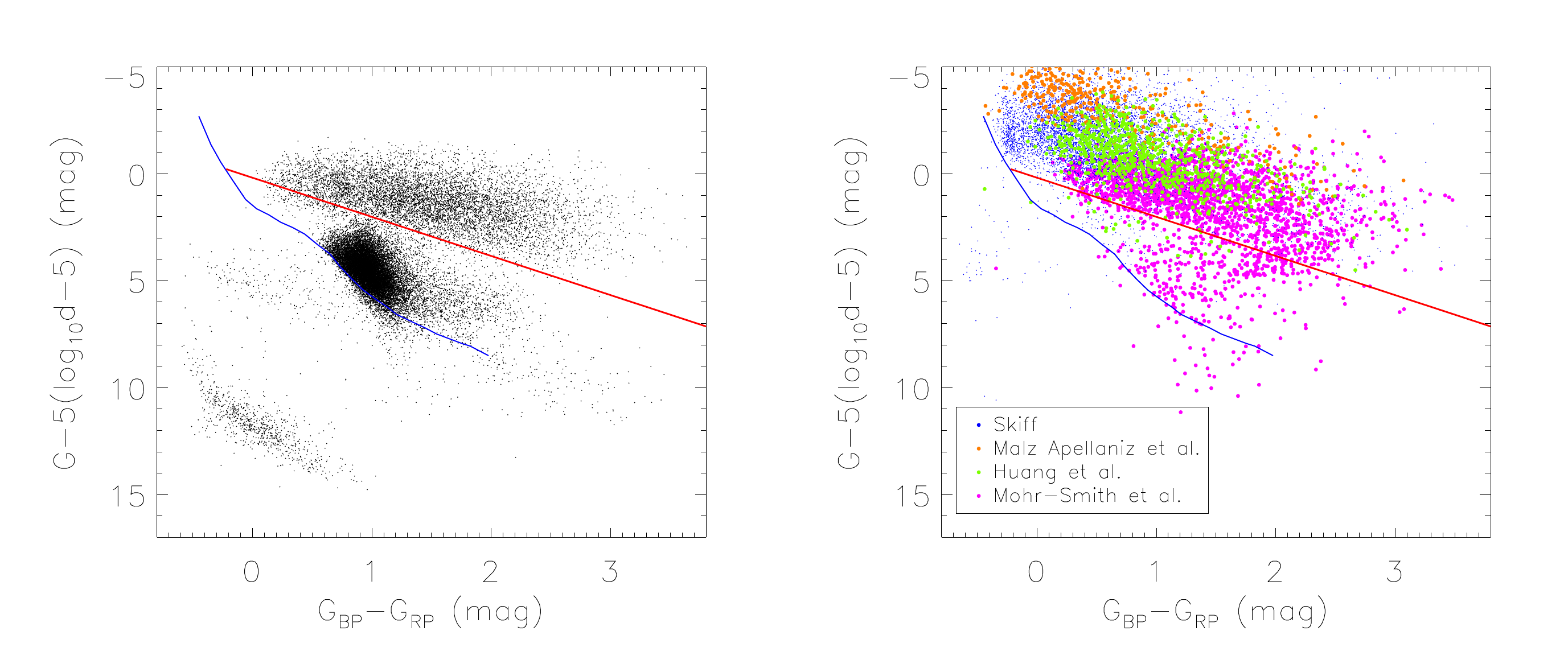}
  \caption{Gaia colour versus `absolute'-magnitude diagrams of 
  OB-star candidates selected from the VPHAS+
  colour-colour diagram (left panel) and those available identified in the literature (right panel). The blue lines 
  represent the PARSEC isochrones \citep{Marigo2017} of main-sequence stars and the red lines 
  are the B3V-star reddening curves. Blue, orange, green and pink dots represent the O-B2  
  candidates from \citet{Skiff2014}, \citet{Maz2013}, Huang et al. (in prep.) and \citet{Mohr2017}, 
 respectively. }
  \label{cmd}
\end{figure*}

  \begin{figure*}
    \centering
  \includegraphics[width=0.75\textwidth]{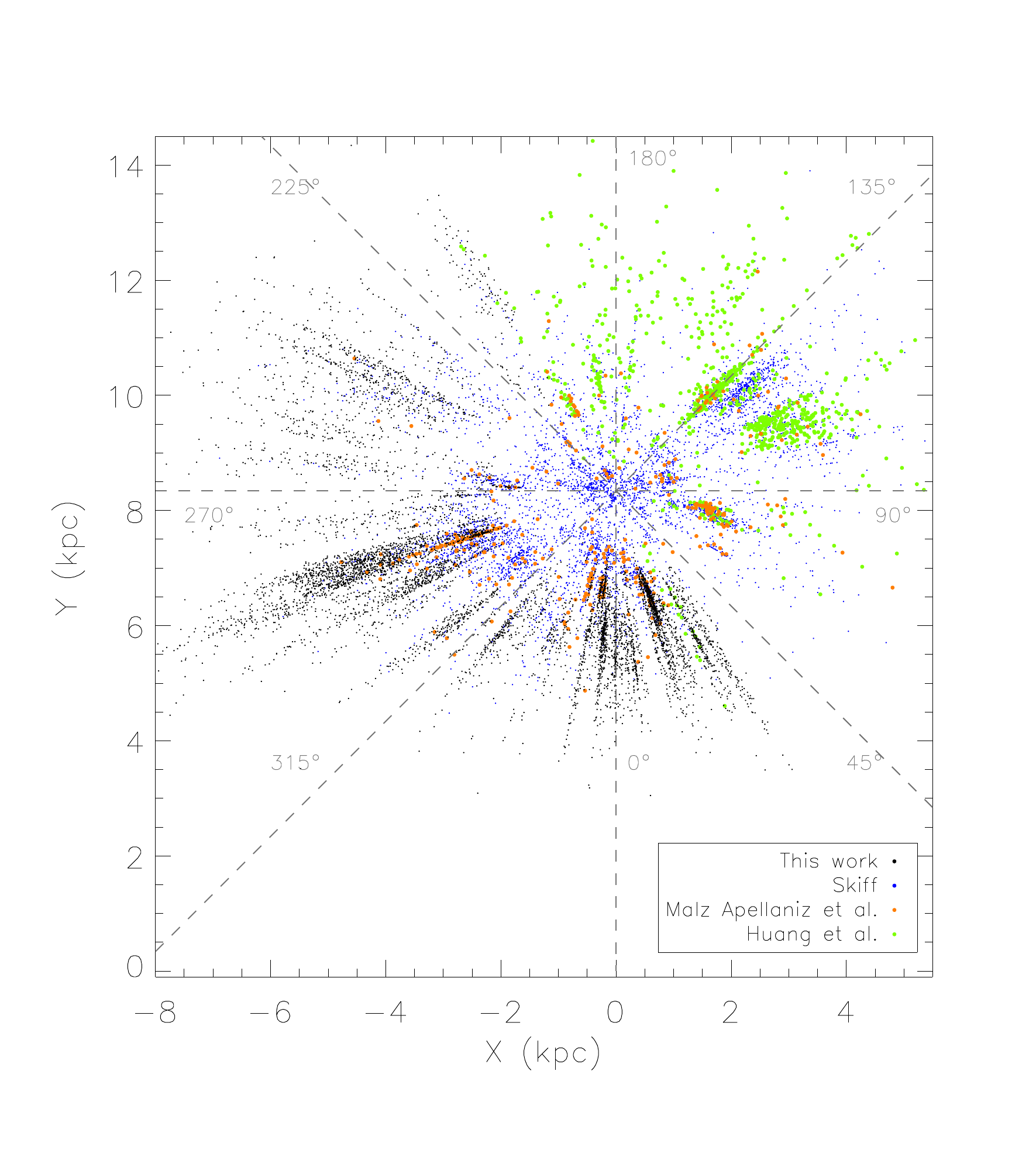}
  \caption{Spatial distribution of the combined OB sample stars in the X–Y plane. 
  Black, blue, orange and green dots correspond to OB star candidates selected in this work, those 
  from  \citet{Skiff2014}, \citet{Maz2013} and Huang et al. (in prep.), respectively.
The Sun, assumed to be at 8.34\,kpc from the Galactic center, is located at the centre
of the plot ($X$ = 0\,kpc and $Y$=8.34\,kpc). 
The directions of $l$ = 0\degr, 45\degr, 90\degr, 135\degr, 180\degr, 225\degr, 
270\degr\ and 315\degr\ are also marked in the plot.}
  \label{xyd}
\end{figure*}

\begin{table*}
   \centering
  \caption{Description of the OB-star candidate catalogue selected from the VPHAS+ DR2.}
  \begin{tabular}{ccl}
  \hline
  \hline
Column & Name & Description  \\
\hline
1 & id & Index of the star in the catalogue  \\
2 & RA &  Right Ascension (J2000)  \\
3 & Dec  & Declination (J2000) \\
4 & u  &  VPHAS+ $u$-band magnitude  \\
5 &  uerr  &  VPHAS+ $u$-band photometric uncertainty   \\
6  & g  & VPHAS+ $g$-band magnitude  \\
7 &  gerr  &  VPHAS+ $g$-band photometric uncertainty  \\
8 &  r &  VPHAS+ $r$-band magnitude  \\
9  & rerr & VPHAS+ $r$-band photometric uncertainty   \\
10 &  i &  VPHAS+ $i$-band magnitude  \\
11 &  ierr & VPHAS+ $i$-band photometric uncertainty   \\
12  & \Ha\ &  VPHAS+ $H\alpha$-band magnitude  \\
13 &  \Ha err &  VPHAS+ $H\alpha$-band photometric uncertainty   \\
14  & G &  Gaia $G$-band magnitude  \\
15  & Gerr &  Gaia $G$-band photometric uncertainty   \\
16  & BP &  Gaia \gb-band magnitude  \\
17 &  BPerr &  Gaia \gb-band photometric uncertainty   \\
18  & RP &  Gaia \gr-band magnitude  \\
19 & RPerr &  Gaia \gr-band photometric uncertainty   \\
20 & dis & Distance of object estimated from Eq.~(1) \\
21 & para & Gaia parallax \\
22 & paraerr & Gaia parallax uncertainty \\
23 & pmra & Gaia proper motion in Right Ascension \\
24 & pmraerr & Uncertainty of proper motion in Right Ascension \\
25 & pmdec & Gaia proper motion in Declination \\
26 & pmdecerr & Uncertainty of proper motion in Declination \\
27 & ref &  The reference if the star is already presented in the literature \\
\hline
\end{tabular} 
\end{table*}

\begin{table}  
\centering
\caption{Numbers and fractions of stars in different spectral types, for a sample of 139 OB star candidates that have spectroscopic
 classifications available from the SIMBAD database.}
\begin{tabular}{lrr}
\hline
\hline
Type & Number & Fraction  \\
\hline
O stars & 19 & 14\,per\,cent    \\
B0-B2 stars & 54 & 39\,per\,cent   \\
B3 stars &  9  &  6\,pre\,cent \\
B4-B8 stars & 13 &  9\,per\,cent  \\
Other OB stars$^1$ & 30 & 22\,per\,cent  \\
WR stars &  5 & 4\,per\,cent \\
A/F/M stars &  9 & 6\,per\,cent    \\
\hline
\end{tabular}
\parbox{75mm}
{\footnotesize \baselineskip 3.8mm
 Notes: $^1$Of OB spectral type but without subclasses available.}
\end{table}

\begin{table*}
   \centering
  \caption{Description of the O-B2 star catalogue compiled from the literature.}
  \begin{tabular}{ccl}
  \hline
  \hline
Column & Name & Description  \\
\hline
1 & RA &  Right Ascension (J2000)  \\
2 & Dec  & Declination (J2000) \\
3  & G &  Gaia $G$-band magnitude  \\
4  & Gerr &  Gaia $G$-band photometric uncertainty   \\
5  & BP &  Gaia \gb-band magnitude  \\
6 &  BPerr &  Gaia \gb-band photometric uncertainty   \\
7  & RP &  Gaia \gr-band magnitude  \\
8 & RPerr &  Gaia \gr-band photometric uncertainty   \\
9 & dis & Distance of object estimated from Eq.~(1) \\
10 & para & Gaia parallax \\
11 & paraerr & Gaia parallax uncertainty \\
12 & pmra & Gaia proper motion in Right Ascension \\
13 & pmraerr & Uncertainty of proper motion in Right Ascension \\
14 & pmdec & Gaia proper motion in Declination \\
15 & pmdecerr & Uncertainty of proper motion in Declination \\
16 & sptype & Spectral type \\
17 & ref &  The reference \\
\hline
\end{tabular} 
\end{table*}

Similarly to \citet{Mohr2015} and \citet{Mohr2017}, we first select OB star candidates 
with the ($u-g$, $g-r$) colour-colour diagram. In the diagram, the reddened OB stars of 
spectral types earlier than B3 are easily identifiable as they are located above and away from the main stellar 
locus. We select sources in the VPHAS+ DR2 catalogue with cuts $g$ $<$ 20\,mag, $g$-band photometric 
errors smaller than 0.05\,mag, $u$ and $r$ detections and photometric uncertainties smaller than 0.05\,mag. 
In Fig.~\ref{ccd}, we plot the distribution of all stars thus selected in the 
($u-g$, $g-r$) colour-colour diagram. 
We adopt the reddening vector for the B3V stars from \citet{Drew2014}. In total,  
91,668 unique stars are found located above the B3V-star reddening curve. 

Our aim is to select a sample of young and luminous OB stars to trace the Galactic spiral arms. 
The candidates selected from the 
 ($u-g$, $g-r$) diagram includes all types of OB stars. One thus needs to identify and 
 reject stars amongst those candidates that actually belong to the old populations, 
 such as sub-dwarfs and white dwarfs. In addition, the sample selected from the VPHAS+
 colour-colour diagram is also contaminated by a significant number of stars of B4- and later-types due to the 
 relative large photometric uncertainties 
 of $u$-band  (see also \citealt{Mohr2017}). To exclude those contaminators, \citet{Mohr2017}
 carried out a SED analysis based on the multi-band photometry of VPHAS+ and 2MASS.
 In the current work, we adopt the high precision photometric data and parallaxes from the Gaia DR2
 to exclude all the contaminators.
 We cross-match the candidates with the Gaia DR2 catalogue.
 We adopt only stars of Gaia parallax errors smaller than 20\,per\,cent,
 re-normalised unit weight errors (RUWE) small than 1.4 (Gaia technical note GAIA-C3-TN-LU-LL-124-01) and
 Gaia simple flux ratios $C = (I_{\rm BP} + I_{\rm RP})/I_G$ smaller than 
 1.3+0.06$(G_{\rm BP}-G_{\rm RP})^2$ \citep{Evans2018}.
This yields 26,936 stars.
We construct a Gaia colour-`absolute' magnitude diagram, without reddening corrections, using 
`absolute' magnitude in Gaia $G$-band, 
 $M_G = G-5({\rm log_{10}}d - 5)$, where $d$ is distance of the star.
The systematic trends and the zero points of the published Gaia $G$ magnitudes in the Gaia DR2 are corrected 
using the relations of \citet{Evans2018} and \citet{Maz2018}\footnote{https://www.cosmos.esa.int/web/gaia/dr2-known-issues}.
The distances of objects in this work are calculated from the Gaia parallaxes using a
Bayesian distance estimator \citep{Maz2005, Bailer2018}, given by,
\begin{equation}
p(d|\varpi) = r^2\exp(-\dfrac{1}{2\sigma^2_{\varpi}}(\varpi-\varpi_{\rm zp} - \dfrac{1}{d}))p(d),
\end{equation}
where $\varpi$ and $\sigma_{\varpi}$ are Gaia parallax and its associated uncertainty,
 $\varpi _ {\rm zp}$ the global parallax zeropoint ($\varpi _ {\rm zp}$ $=$ $-$0.029  from \citealt{Lindegren2018}) and 
 $p(d)$ the space density distribution prior for the Galactic OB stars adopted from \citet{Maz2008}.

The Gaia colour-`absolute' magnitude diagram for the 26,936 selected candidates in 
the current work is shown in Fig.~\ref{cmd}. The stars fall clearly 
into four groups, a group of  young and luminous OB stars located at the top of the diagram with 
$M_G$ $\sim$ 0\,mag,  a second group of white dwarfs of $M_G$ $>$ 10\,mag, a third group of sub-dwarfs of 
$M_G$ $>$ 2\,mag and $(G_{\rm BP} - G_{\rm RP})$ $<$ 1\,mag (below the PARSEC isochrone) and a fourth group of B4- and later-type stars of 
$M_G$ $>$ 2\,mag and $(G_{\rm BP} - G_{\rm RP})$ $>$ 1\,mag (above the PARSEC isochrone). 
The extinction vector for a B3V star, calculated with the extinction law of \citet{Maz2014} assuming   
$R_{5495}$ = 3.1 \citep{Maz2018b}, nicely separates the OB stars and the contaminators.
Most O-B2 candidates from the literature are
located well above the B3V extinction curve.
Table~1 gives the number of stars selected in each step of our OB-star selection procedure. 
The majority of the sample in Fig.~\ref{cmd} are B4- and later-type stars (17,906 out of 26,936). This is largely 
due to the relatively large uncertainties of the VPHAS+ $u$ band magnitudes. 
  
Based on the above exercise, 6,858 OB candidates have been selected 
above the B3V-star extinction curve. A catalogue of 
these OB-star candidates is available in electronic form in the online version of this 
manuscript\footnote{The catalogue is also available online via \\
``http://paperdata.china-vo.org/diskec/obsamp/table2.fits''.}. Table~2 describes
the data format of the catalogue. In order to check if any objects in our catalogue
have known spectral types, we cross-match our candidates with the SIMBAD 
database with a match radius of 1\,arcsec. We find 139 objects that have 
been spectroscopically classified in the SIMBAD.
Table~3 summarises the numbers and fractions of those stars in different spectroscopically confirmed spectral types.  
The majority of them are indeed OB stars as expected, with a contamination at the level of 
$\sim$ 10\,pre\,cent from Wolf-Rayet (WR) stars, and later A, F and M stars. 
A very positive feature of Table~3 is the high success-rate that we are able to place stars correctly into
the O and early B spectral types. Most of the OB stars are of early types (B3 and earlier), 
with only about 9\,per\,cent contamination of B4 and later types. 
We have also checked our OB-star candidates against data from large-scale spectroscopic surveys, and find 
27 common stars with the LAMOST surveys. 
There are 25 stars that have effective temperatures higher than $\sim$ 12,000\,K in the value-added catalogue of the LAMOST
Spectroscopic Survey of the Galactic Anti-centre Data Release 2 
(LSS-GAC DR2; \citealt{Xiang2017}). 

\subsection{OB stars from the literature}

In addition to the new OB candidates selected from the VPHAS+ DR2 catalogue, we have also made use of the 
OB stars available in the literature. In the current work, we adopt three catalogues, the 
GOSC catalogue \citep{Maz2013}, the OB star catalogue selected from \citet{Skiff2014} and 
that of Huang et al. (in prep.) selected from 
the LAMOST Spectroscopic Survey of the Galactic Anti-centre (LSS-GAC; \citealt{Liu2014,Yuan2015}). 

\citet{Skiff2014} present a catalogue of $\sim$ 0.9 million stars 
with spectral classifications complied from the literature. In this work, we select O-B2 stars from this catalogue 
and cross-match them with the Gaia DR2 catalogue. After excluding stars 
with parallax uncertainties larger than 20\,per\,cent, RUWEs larger than 1.4 and 
$C = (I_{\rm BP} + I_{\rm RP})/I_G$ larger than 
 1.3+0.06$(G_{\rm BP}-G_{\rm RP})^2$, we are
left with 7,441 unique objects. Amongst those, 
93 objects are common with the new VPHAS+ sample.

The objects of the GOSC catalogue are identified from the 
new, homogeneous, high signal-to-noise ratio, and low spectral resolution GOSSS survey.
The current version of the catalogue (v4.1) contains 594 O stars, 24 BA stars, and 11 late-type stars. 
We select O-B2 stars from the catalogue 
and cross-match them with the Gaia DR2 catalogue. 
Again, stars with parallax uncertainties larger than 20\,per\,cent, RUWEs larger than 1.4 and 
$C = (I_{\rm BP} + I_{\rm RP})/I_G$ larger than 
 1.3+0.06$(G_{\rm BP}-G_{\rm RP})^2$ are
excluded. This yields 442 unique objects. 
There are 246 common objects between the GOSC and the Skiff samples. Only 2 common
objects are found between the GOSC and the new VPHAS+ samples.

The LAMOST telescope has surveyed several millions stars in  
a contiguous area of the Galactic disk toward the Galactic anti-centre (150\degr\ $<$ $l$ $<$ 210\degr and 
$|b|$ $<$ 30\degr). From the LAMOST spectra, 
Huang et al. (in prep.) have identified $\sim$25,000 OB stars with Gaia parallax errors smaller than 20\,per\,cent,
RUWEs smaller than 1.4 and 
$C = (I_{\rm BP} + I_{\rm RP})/I_G$ smaller than 
 1.3+0.06$(G_{\rm BP}-G_{\rm RP})^2$.
931 of them are identified as O-B2 stars. There are 237 common stars between the Huang et al and the Skiff samples.
Only 6 common stars are found between the Huang et al and the GOSC samples.
As the LAMOST surveys are spectroscopic and the survey areas
do not have much overlap with the VPHAS+ survey, there are only 3 common stars between the 
Huang et al. sample and the new VPHAS+ sample.

This catalogue of all O-B2 stars compiled from the literature is available in electronic 
form in the online version of this manuscript\footnote{The catalogue is also available online via \\
``http://paperdata.china-vo.org/diskec/obsamp/table4.fits''.}. Table~4 describes the format of the catalogue. 

\subsection{The combined sample of OB  stars and candidates}

Combing our newly selected VPHAS+ OB-star candidates with those from the literature, we obtain a catalogue of 
14,880 unique OB stars and candidates, all with Gaia parallax uncertainties smaller than 20\,per\,cent. 
Distances of all stars in the catalogue are calculated from the Gaia parallaxes using Eq.~(1).
We have compared our distance estimates with those of \citet{Bailer2018} and found no systematics 
between them and the dispersion is very small (about 1\,pre\,cent). 
In Fig.~\ref{xyd}, we show the distribution of those OB stars and candidates in the Galactic $X$-$Y$ plane.  This new 
combined sample spans from $X$ = $-$8 to 5\,kpc and $Y$ = 2 to 15\,kpc in the Galactic plane. 
The VPHAS+ and the literature samples complement with each other
in the spatial coverage. The Skiff and Ma{\'{\i}}z Apell{\'a}niz et al. sample stars are mainly distributed 
at distances $d$ $<$ 4\,kpc from the Sun, while 
those of Huang et al. and those newly selected from the VPHAS+ fall between 
 2 $<$  $d$ $<$ 6\,kpc from the Sun. Stars from Huang et al. are mainly in the Galactic outer disk while 
 those of the VPHAS+ are in the Galactic inner disk.

\section{The Galactic spiral arms}

 \begin{figure*}
    \centering
  \includegraphics[width=0.7\textwidth]{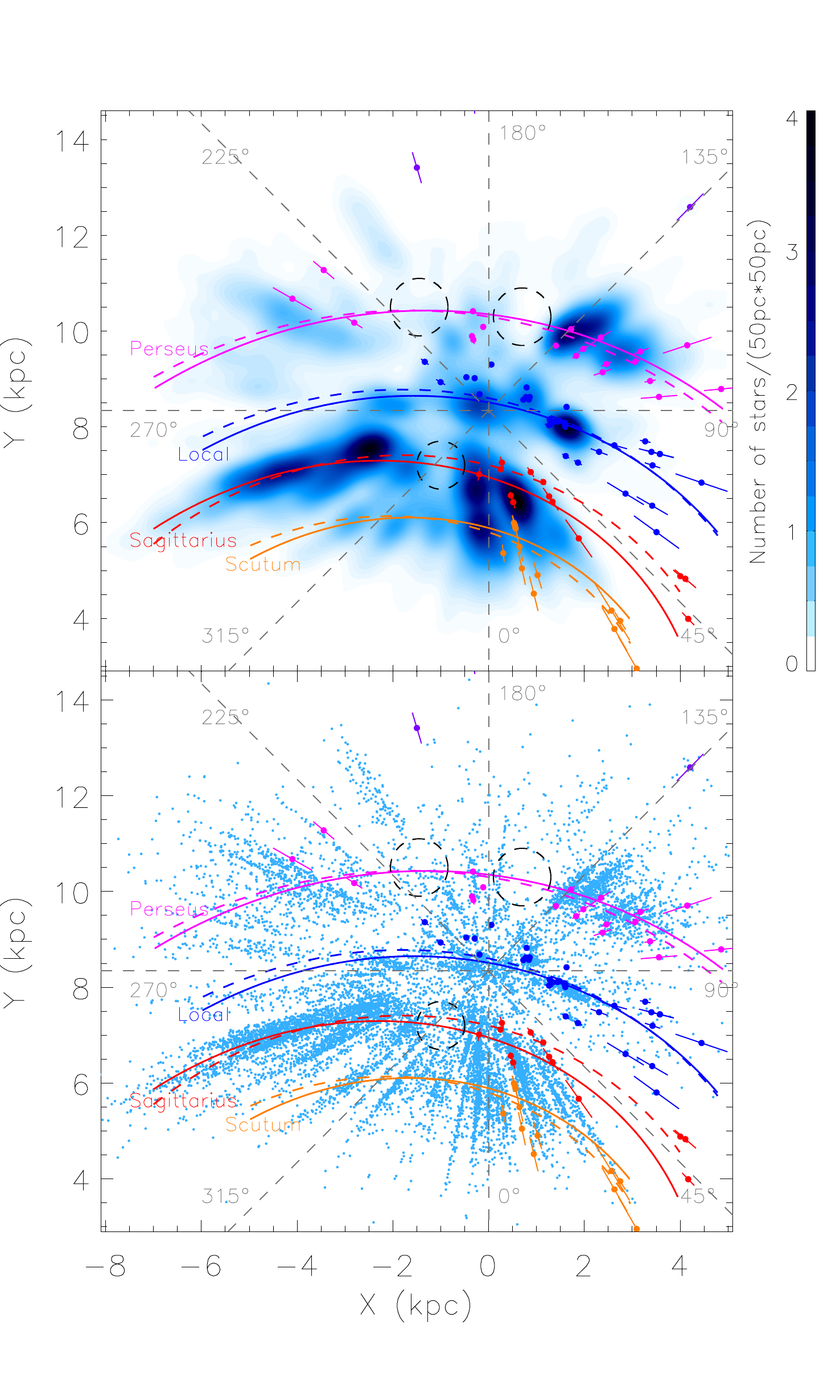}
  \caption{Spatial (bottom panel) and  
  density (upper panel) distributions 
  of the combined sample of OB stars and candidates in the $X$-$Y$ plane. 
Solid and dashed orange, red, blue and pink lines delineate respectively 
  the best-fit spiral arm models of the Scutum, Sagittarius, Local and Perseus Arms presented in the current work and those of
  \citet{Xu2018b}. Circles of the above colours are the masers 
  from \citet{Xu2018b} probably associated with the individual Arms.
The purple circles are the masers from \citet{Xu2018b} that may trace the Outer Arm.   
Three black dashed circles mark the positions of possible `hole' patterns in the 
Sagittarius and Perseus Arms.
The Sun, assumed to be at 8.34\,kpc from the Galactic center, is located at the centre
of the plot. The directions of $l$ = 0\degr, 45\degr, 90\degr, 135\degr, 180\degr, 225\degr, 
270\degr\ and 315\degr\ are also marked in the plot.}
  \label{arm}
\end{figure*}

 \begin{figure*}
    \centering
  \includegraphics[width=0.75\textwidth]{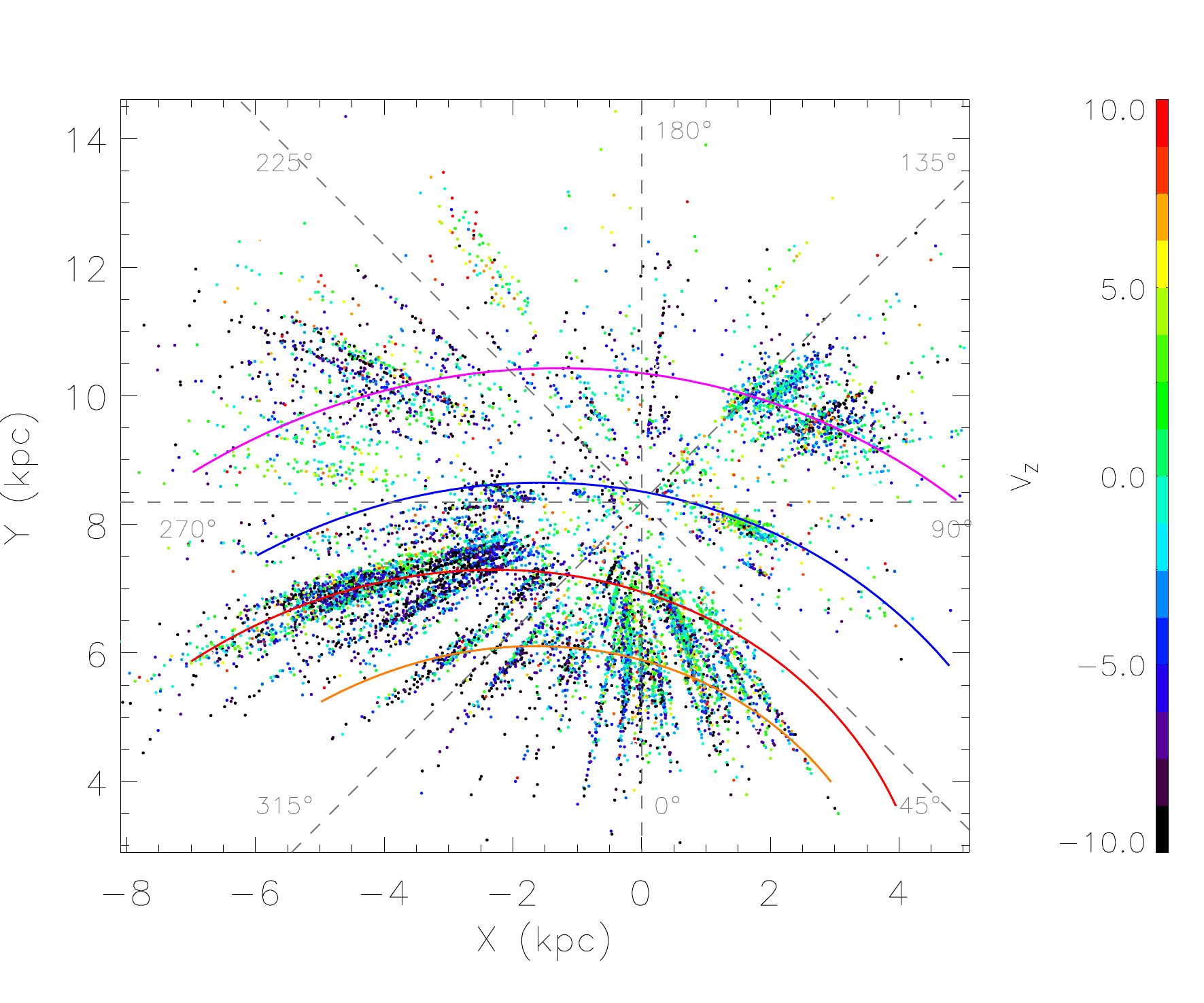}
  \caption{Distribution of values of vertical velocity $V_Z$ for 11,437 OB-star candidates of Galactic latitude $|b|$ $<$ 2\degr.
Orange, red, blue and pink solid lines are respectively 
    the best-fit spiral arm models of the Scutum, Sagittarius, Local and Perseus Arms presented in the current work.
The Sun, assumed to be at 8.34\,kpc from the Galactic center, is located at the centre
of the plot. The directions of $l$ = 0\degr, 45\degr, 90\degr, 135\degr, 180\degr, 225\degr, 
270\degr\ and 315\degr\ are also marked in the plot.
  }
  \label{vz}
\end{figure*}

\subsection{Morphology of the Galactic spiral structure}

As Fig.~\ref{xyd} shows, the OB stars fall in clumps and strips, 
and trace clearly the structure of the Galactic spiral arms. 
The gaps between the arms are quite visible. Comparing with previous 
work \citep{Reid2014, Xu2018, Xu2018b}, our data have a larger spatial coverage 
and probe much further distances from the Sun. Thanks to the
deep limiting magnitudes of the VPHAS+ and LAMOST data, we are now able to 
explore the disk area of distances between 4 and 6\,kpc from the Sun for the first time.

In Fig.~\ref{arm}, we plot the spatial and density distributions of the OB stars in the $X$-$Y$
plane. The density is calculated using a Kernel Density Estimation (KDE) of a Gaussian Kernel of bandwidth 0.2\,kpc.
The masers from the literature (Table~2 of \citealt{Xu2018b}) are over-plotted in the diagram.  
The similarity in the distributions of these two types of tracers is clearly visible. 
This is not surprising as masers and young OB associations   
are both good tracers of massive star forming regions.  
Overall, the spiral pattern revealed by the OB stars and the masers consists of four spiral arm segments. 

In the top parts of the two panels of Fig.~\ref{arm} (Galactocentric distance $R$ $>$ 12\,kpc), there are two masers (purple circles), 
probably the sign posts of the position of the Outer Arm. Limited by the Gaia parallax 
uncertainties, only a few OB stars are found in this region in the current work. Nevertheless, several OB 
stars are indeed found around one of the two masers ($l$ $\sim$ 135\degr).  Two strips of OB stars, one 
at ($X,~Y$) $\sim$ (2.5\,kpc, 12\,kpc) and another  at  ($X,~Y$) $\sim$ ($-$2\,kpc, 12\,kpc), are visible, and they
could belong to the Outer Arm, or to the bridge that connects the Outer and the Perseus Arms.  

The Perseus Arm is well constrained by both the OB stars and 
the masers from $l$ = 90\degr\ to 270\degr. 
In the regions of $l$ between  135\degr\ and 170\degr\ and $l$ between  190\degr\ and 220\degr,
there are two `holes' without masers or few OB stars. The holes are probably true as 
these regions are well covered by the VPHAS+ and LAMOST data. OB stars are clearly found  
in the same directions at
either nearer or further distances from the Sun. Thus it is unlikely that we miss 
the OB stars in the holes should they exist. It is interesting to note that 
the two possible bridges that connect the Outer and the Perseus Arms fall
in those two directions. On the near side of the two holes
($X$ = $-$1.5 to 1.5\,kpc and $Y$ $\sim$ 9\,kpc), several groups of
OB stars as well as some masers are found between the Perseus and the Local Arms.    

The Local Arm is the nearest spiral arm to our Sun and the most well defined in the plot.  
All the OB clumps (or strips) of the other arms show significant linear patterns (``fingers'') that point toward the 
Sun, which are possibly the results of the discontinuous distribution of the VPHAS+ and LAMOST fields,
and have relatively large distance dispersions of about 1\,kpc,
which are likely mainly caused by the relatively large errors in the distances. The Local Arm are well resolved 
into several small OB clumps, with our Sun located near one of them. The presence of OB stars near 
($X,~Y$) = ($-$6\,kpc, 7.5\,kpc) suggests that the Local Arm may extend into the fourth quadrant at $l$ $\sim$ 280\degr, 
indicating that the Local Arm might well be a major spiral arm, rather than a spur structure.

The Sagittarius Arm is well constrained by the OB clumps from $l$ = 290\degr\ ($X$ = $-$7\,kpc
and $Y$ = 5.5\,kpc) to 30\degr\ ($X$ = 2\,kpc and $Y$ = 6\,kpc). A possible `hole' pattern is 
visible in the directions of $l$ between 320\degr\ and 350\degr. Similar to the `holes' in the Perseus Arm, 
there are several OB clumps found at both the nearer and further sides of the `hole', connecting the 
Sagittarius Arm with the Local and the Scutum Arms, respectively. There are a lot more OB stars falling along the 
Sagittarius Arm than those on the Local and the Perseus Arms, suggesting that the star-forming 
activities are much stronger in the inner disk than those in the outer disk.   

The gap between the Scutum and the Sagittarius Arms is not significant in the first quadrant. Due to the large 
distance uncertainties of stars in that region, we are not able to distinguish whether this small gap 
is largely caused by the effects of the large distance errors 
or the two Arms actually merge into one in this region. The gap between the two Arms in the fourth 
quadrant is clearly visible. 

In general, all the Spiral Arms discussed above are well traced by the discrete OB clumps and masers.  In addition to `holes',  
many more possible patterns, such as branches, spurs and bridges between the major spiral arms, are identifiable. 
We therefore can confirm the conclusion of \citet{Xu2018b} that the spiral structure of our Galaxy as traced by the young OB stars
is flocculent, instead of a pure, grand-design spiral structure.

\begin{table}
   \centering
  \caption{Spiral arm characteristics}
  \begin{tabular}{lcccc}
  \hline
  \hline
Arm & $R_0$ & $\psi$ &  $\langle V_Z \rangle$ & $\sigma_{V_Z}$  \\
  &   (kpc) & (\degr) & \kms\ & \kms\ \\ 
\hline
Scutum Arm &  5.89$\pm$0.02 &  15.1$\pm$0.7 & $-$3.5 & 6.5 \\
Sagittarius Arm &  6.95$\pm$0.01 &  17.4$\pm$0.2  & $-$4.1 & 6.8 \\
Local Arm &  8.51$\pm$0.01 &  10.2$\pm$0.2  & $-$2.0 & 7.1 \\
Perseus Arm &  10.35$\pm$0.01 &  7.0$\pm$0.3  & $-$3.9& 6.9 \\
\hline
\end{tabular} 
\end{table}

\subsection{Spiral arm models}

Following \citet{Hou2014} and \citet{Xu2018b}, we derive the parameters of the arm structures, including
the Scutum, Sagittarius, Local and Perseus Arms, based on the current sample of OB stars candidates, as well as masers.  
We adopt the simple, logarithmic model of a spiral arm \citep{Kennicutt1981}, i.e.,
\begin{equation}  
\ln(R/R_0)=-\beta\tan\psi,
\end{equation}
where $R$ is the Galactocentric distance, $\beta$ the Galactocentric azimuth that
has a value of 0 in the direction toward the Sun and increases clockwise, $R_0$ the radius of the arm
at reference azimuth $\beta_0$ = 0\degr\ and $\psi$ the pitch angle of the arm. 
We perform an MCMC analysis to find the optimized value of each of the parameters by
minimising a factor defined as,
\begin{equation}
Z=\dfrac{\sum W_i \sqrt{(x_{i}-x_{t})^{2}+(y_{i}-y_{t})^{2}}}{\sum W_i},
\end{equation}
where $W$ is the weight, $x$ and $y$ 
the Cartesian coordinates, $i$ and $t$ the indexes of tracers and model positions closest 
to the tracers. In the current work, we simply adopt
a weight $W_{\rm OB}$ = 1 for all the OB stars and candidates and 
$W_{\rm m} $ = 10 for all the masers \citep{Xu2018b}.
The errors of the derived parameters
are calculated using the bootstrap method \citep{Wall2003}. 
We randomly re-sampled the combined OB sample 1000 times 
and obtain the best-fit parameters for each sample. 
The r.m.s scatters of the resulted parameters
give the uncertainties of our results.

The best-fitted models of the individual arms are plotted in Fig.~\ref{arm} and the corresponding 
parameters are listed in Table~5. Our models connect most of the OB clumps and masers.
The resulted values of distance $R_0$ of the four spiral arms are very similar to those from \citet{Xu2018b}, who
obtain the structure parameters of the four arms using O stars from \citet{Reed2003} and the masers. 
Our current sample of traces covers a larger range in the $X$-$Y$ space than that of \citet{Xu2018b}. And this yields the pitch angles of
the individual arms that differ slightly from those of \citet{Xu2018b} but match better with the new data. 

\subsection{Vertical motions of the spiral arms}

The Gaia DR2 catalogue has provided us accurate proper motions for the OB stars in our catalogue. As most of 
the OB stars are located in the Galactic plane of $b$ $\sim$ 0\degr, the vertical components of their
radial velocities $V_R$ can be ignored. Thus the vertical velocities of 
the individual targets 
in our catalogue can be simply calculated from Gaia proper motions. We select 11,437 OB stars and candidates of Galactic latitude $|b|$ 
$<$ 2\degr\ and calculate values of their vertical velocity $V_Z$. The distribution of values thus derived is 
plotted in Fig.~\ref{vz}. Similar to the spatial position distribution, the $V_Z$ distribution of these OB stars
are also striped and clumped. We have estimated the means and dispersions of $V_Z$ of the individual 
spiral arms and the results are listed in Table~5. In general, the spiral arms have a negative mean 
vertical motion that varies slightly from one arm to another. The dispersions of $V_Z$ of the 
individual arms  are quite similar,  $\sim$ 7\,\kms.

\section{Discussion and summary}

 \begin{figure}
    \centering
  \includegraphics[width=0.49\textwidth]{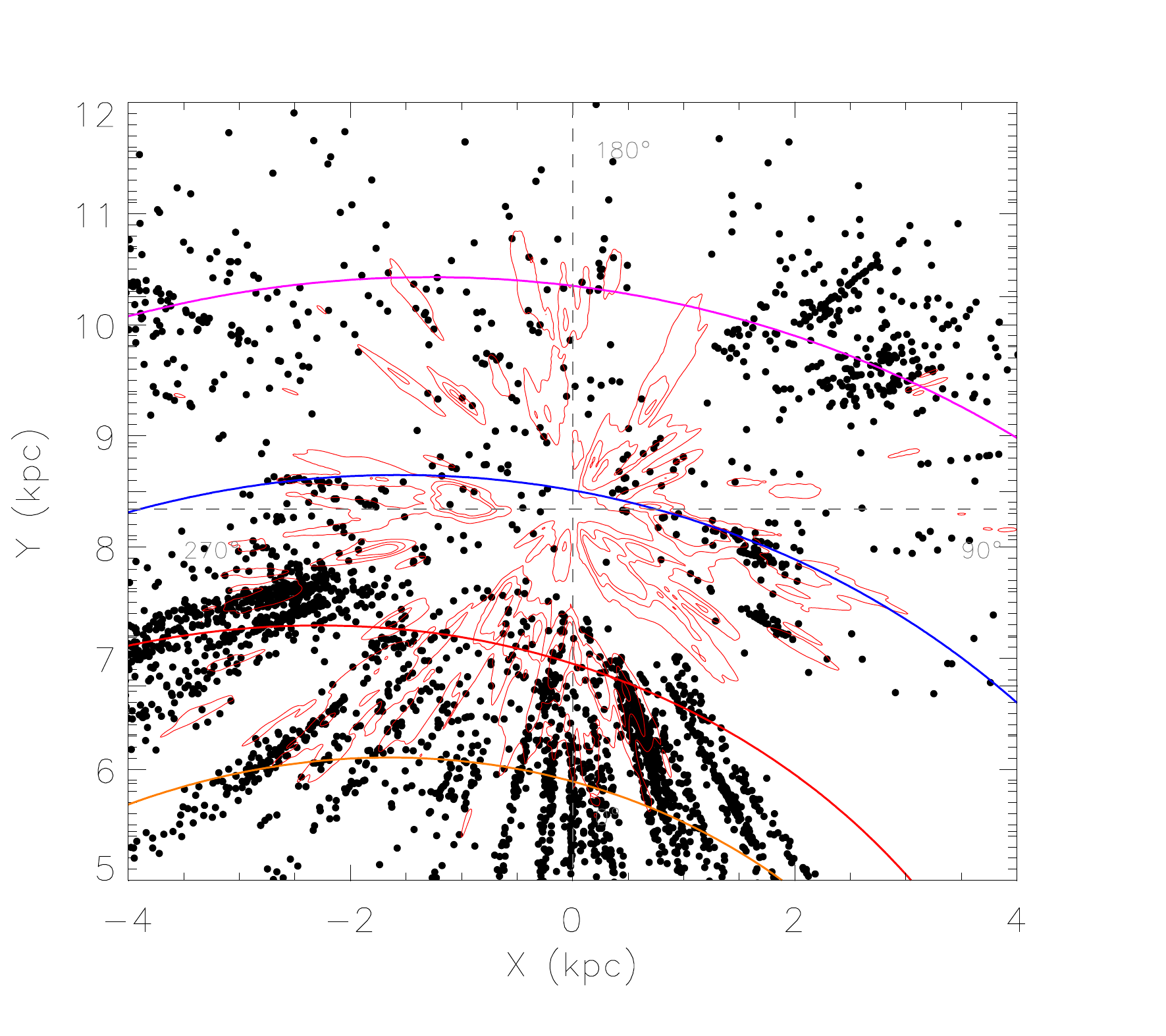}
  \caption{The spatial distribution of stars of $|b|$ $<$ 0.5\degr\ in the combined sample of OB stars and candidates 
   (black points) compared to that of the 
 interstellar dust reddening of $|b|$ $<$ 0.5\degr\ (\citealt{Chen2018}, red contours).
Orange, red, blue and pink solid lines are respectively the 
   best-fit spiral arm models of the Scutum, Sagittarius, Local and Perseus Arms.
The red contours correspond to the differential values of \ebr\ of the local dust grains that range from 0 to 12\,mag\,kpc$^{-1}$.
The Sun, assumed to be at 8.34\,kpc from the Galactic center, is located at the centre
of the plot. The directions of $l$ = 0\degr, 90\degr, 180\degr\ and 270\degr\ are also 
marked in the plot.}
  \label{ebr}
\end{figure}

In Fig.~\ref{ebr} we compare the spatial distribution of our OB stars and candidates with that of 
the interstellar dust reddening for $|b|$ $<$ 0.5\degr\ of the Galactic plane \citep{Chen2018}. 
Due to the heavy extinction in the Galactic plane, the extinction map of \citet{Chen2018}
is only completed to $\sim$ 3\,kpc.
Overall, there is a good correlation between the distribution of the OB stars and candidates and that of the
interstellar dust at large scales. The Sagittarius, Local and Perseus arms are discernible in both
the interstellar dust extinction map and in the OB star distribution. The dust clouds are very likely
to be spatially associated with the Galactic spiral arm models delineated
in the current work. 

In this paper, we have identified 6,858 new O- and early B-type star candidates from the VPHAS+ DR2 and 
Gaia DR2 catalogues. Combined with the O-B2 candidates available in the literature, we have built
a sample of 14,880 O- and early B-type stars and candidates with Gaia parallax errors smaller than 20\,per\,cent. 
This is  hitherto the largest sample of O- and early B-type stars and candidates with accurate distance and proper motion estimates. 
Based on the catalogue, we have explored the morphology and kinematics of the Galactic
spiral structure. Our sample reveal clearly four spiral arm segments, i.e., the Scutum,   
Sagittarius, Local and Perseus Arms. We have obtained accurate structure
parameters of those Arms. 
The data show three possible `hole' patterns along the Galactic spiral arms, in addition to abundant 
other substructures.  The Galactic spiral structures as traced by the young OB stars are more likely
flocculent spirals.

The size and the spatial distribution of our sample are mainly limited by the parallax uncertainties of  the
Gaia DR2. If we loose the Gaia parallax uncertainties up to 30\,per\,cent, the number of the 
selected VPHAS+ OB candidates would be doubled and the sample will extend to
a distance of $\sim$ 10\,kpc from the Sun.
However, the increased distance errors will lead to a larger fraction of contamination of the sub-dwarfs. Meanwhile the  
distance uncertainty might become comparable to or even larger than the gaps between the spiral arms. 
The future Gaia releases are expected to improve our work by enlarging the sample size and also the spatial coverage. 
On the other hand, VPHAS+ DR2, that we used in the current work, covers only 
20\,per\,cent of the full footprint of the VPHAS+ survey. 
The future VPHAS+ data release would also help enlarge our data set and extend to higher Galactic latitudes. 
Other ongoing and future $u$-band photometric surveys,
such as the SkyMapper, the Large Synoptic Survey Telescope (LSST) and 
the Multi-channel Photometric Survey Telescope (Mephisto) surveys, will aid in the construction of samples of more 
O- and early B-type stars that cover larger areas of the Galactic plane.

\section*{Acknowledgements}

We want to thank our anonymous referee for the insightful
comments. This work is partially supported by National
Natural Science Foundation of China 11803029, U1531244, 11833006 and U1731308
and Yunnan University grant No.~C176220100007. 
LGH is supported by the National Key R\&D Program of China (NO. 2017YFA0402701) and the Youth Innovation Promotion Association CAS.
HBY is supported by NSFC grant  
No.~11603002 and Beijing Normal University grant No.~310232102.
This research made use of the cross-match service provided by CDS, Strasbourg.

This work has made use of data products from the Guoshoujing Telescope (the
Large Sky Area Multi-Object Fibre Spectroscopic Telescope, LAMOST). LAMOST
is a National Major Scientific Project built by the Chinese Academy of
Sciences. Funding for the project has been provided by the National
Development and Reform Commission. LAMOST is operated and managed by the
National Astronomical Observatories, Chinese Academy of Sciences.

Based on data products from observations made with ESO Telescopes at the La Silla Paranal Observatory under programme ID 177.D-3023, as part of the VST Photometric H$\alpha$ Survey of the Southern Galactic Plane and Bulge (VPHAS+, www.vphas.eu).

This work presents results from the European Space Agency (ESA) space mission Gaia. Gaia data are being processed by the Gaia Data Processing and Analysis Consortium (DPAC). Funding for the DPAC is provided by national institutions, in particular the institutions participating in the Gaia MultiLateral Agreement (MLA). The Gaia mission website is https://www.cosmos.esa.int/gaia. The Gaia archive website is https://archives.esac.esa.int/gaia.

\bibliographystyle{mn2e}
\bibliography{obsamp}

\begin{thebibliography}{51}
\expandafter\ifx\csname natexlab\endcsname\relax\def\natexlab#1{#1}\fi

\bibitem[{{Bailer-Jones} {et~al.}(2018){Bailer-Jones}, {Rybizki}, {Fouesneau},
  {Mantelet}, \& {Andrae}}]{Bailer2018}
{Bailer-Jones}, C.~A.~L., {Rybizki}, J., {Fouesneau}, M., {Mantelet}, G., \&
  {Andrae}, R. 2018, \aj, 156, 58

\bibitem[{{Caswell} \& {Haynes}(1987)}]{Caswell1987}
{Caswell}, J.~L. \& {Haynes}, R.~F. 1987, \aap, 171, 261

\bibitem[{{Chen} {et~al.}(2019){Chen}, {Huang}, {Yuan}, {Wang}, {Fan}, {Xiang},
  {Zhang}, {Tian}, \& {Liu}}]{Chen2018}
{Chen}, B.-Q., {et~al.} 2019, \mnras, 483, 4276

\bibitem[{{Dame} \& {Thaddeus}(2011)}]{Dame2011}
{Dame}, T.~M. \& {Thaddeus}, P. 2011, \apjl, 734, L24

\bibitem[{{Deng} {et~al.}(2012){Deng}, {Newberg}, {Liu}, {Carlin}, {Beers},
  {Chen}, {Chen}, {Christlieb}, {Grillmair}, {Guhathakurta}, {Han}, {Hou},
  {Lee}, {L{\'e}pine}, {Li}, {Liu}, {Pan}, {Sellwood}, {Wang}, {Wang}, {Yang},
  {Yanny}, {Zhang}, {Zhang}, {Zheng}, \& {Zhu}}]{Deng2012}
{Deng}, L.-C., {et~al.} 2012, Research in Astronomy and Astrophysics, 12, 735

\bibitem[{{Downes} {et~al.}(1980){Downes}, {Wilson}, {Bieging}, \&
  {Wink}}]{Downes1980}
{Downes}, D., {Wilson}, T.~L., {Bieging}, J., \& {Wink}, J. 1980, \aaps, 40,
  379

\bibitem[{{Drew} {et~al.}(2014){Drew}, {Gonzalez-Solares}, {Greimel}, {Irwin},
  {K{\"u}pc{\"u} Yoldas}, {Lewis}, {Barentsen}, {Eisl{\"o}ffel}, {Farnhill},
  {Martin}, {Walsh}, {Walton}, {Mohr-Smith}, {Raddi}, {Sale}, {Wright},
  {Groot}, {Barlow}, {Corradi}, {Drake}, {Fabregat}, {Frew}, {G{\"a}nsicke},
  {Knigge}, {Mampaso}, {Morris}, {Naylor}, {Parker}, {Phillipps}, {Ruhland},
  {Steeghs}, {Unruh}, {Vink}, {Wesson}, \& {Zijlstra}}]{Drew2014}
{Drew}, J.~E., {et~al.} 2014, \mnras, 440, 2036

\bibitem[{{Evans} {et~al.}(2018){Evans}, {Riello}, {De Angeli}, {Carrasco},
  {Montegriffo}, {Fabricius}, {Jordi}, {Palaversa}, {Diener}, {Busso},
  {Cacciari}, {van Leeuwen}, {Burgess}, {Davidson}, {Harrison}, {Hodgkin},
  {Pancino}, {Richards}, {Altavilla}, {Balaguer-N{\'u}{\~n}ez}, {Barstow},
  {Bellazzini}, {Brown}, {Castellani}, {Cocozza}, {De Luise}, {Delgado},
  {Ducourant}, {Galleti}, {Gilmore}, {Giuffrida}, {Holl}, {Kewley}, {Koposov},
  {Marinoni}, {Marrese}, {Osborne}, {Piersimoni}, {Portell}, {Pulone},
  {Ragaini}, {Sanna}, {Terrett}, {Walton}, {Wevers}, \&
  {Wyrzykowski}}]{Evans2018}
{Evans}, D.~W., {et~al.} 2018, \aap, 616, A4

\bibitem[{{Gaia Collaboration} {et~al.}(2018){Gaia Collaboration}, {Brown},
  {Vallenari}, {Prusti}, {de Bruijne}, {Babusiaux}, {Bailer-Jones}, {Biermann},
  {Evans}, {Eyer}, \& et~al.}]{Gaia2018}
{Gaia Collaboration}, {et~al.} 2018, \aap, 616, A1

\bibitem[{{Georgelin} \& {Georgelin}(1976)}]{Georgelin1976}
{Georgelin}, Y.~M. \& {Georgelin}, Y.~P. 1976, \aap, 49, 57

\bibitem[{{Heyer} {et~al.}(2001){Heyer}, {Carpenter}, \& {Snell}}]{Heyer2001}
{Heyer}, M.~H., {Carpenter}, J.~M., \& {Snell}, R.~L. 2001, \apj, 551, 852

\bibitem[{{Honma} {et~al.}(2007){Honma}, {Bushimata}, {Choi}, {Hirota}, {Imai},
  {Iwadate}, {Jike}, {Kameya}, {Kamohara}, {Kan-Ya}, {Kawaguchi}, {Kijima},
  {Kobayashi}, {Kuji}, {Kurayama}, {Manabe}, {Miyaji}, {Nagayama}, {Nakagawa},
  {Oh}, {Omodaka}, {Oyama}, {Sakai}, {Sato}, {Sasao}, {Shibata}, {Shintani},
  {Suda}, {Tamura}, {Tsushima}, \& {Yamashita Kazuyoshi}}]{Honma2007}
{Honma}, M., {et~al.} 2007, \pasj, 59, 889

\bibitem[{{Hou} \& {Han}(2014)}]{Hou2014}
{Hou}, L.~G. \& {Han}, J.~L. 2014, \aap, 569, A125

\bibitem[{{Hou} {et~al.}(2009){Hou}, {Han}, \& {Shi}}]{Hou2009}
{Hou}, L.~G., {Han}, J.~L., \& {Shi}, W.~B. 2009, \aap, 499, 473

\bibitem[{{Johnson} \& {Morgan}(1953)}]{Johnson1953}
{Johnson}, H.~L. \& {Morgan}, W.~W. 1953, \apj, 117, 313

\bibitem[{{Kalberla} \& {Kerp}(2009)}]{Kalberla2009}
{Kalberla}, P.~M.~W. \& {Kerp}, J. 2009, \araa, 47, 27

\bibitem[{{Kennicutt}(1981)}]{Kennicutt1981}
{Kennicutt}, Jr., R.~C. 1981, \aj, 86, 1847

\bibitem[{{Kerr}(1969)}]{Kerr1969}
{Kerr}, F.~J. 1969, \araa, 7, 39

\bibitem[{{Koo} {et~al.}(2017){Koo}, {Park}, {Kim}, {Lee}, {Balser}, \&
  {Wenger}}]{Koo2017}
{Koo}, B.-C., {Park}, G., {Kim}, W.-T., {Lee}, M.~G., {Balser}, D.~S., \&
  {Wenger}, T.~V. 2017, \pasp, 129, 094102

\bibitem[{{Kuijken}(2011)}]{Kuijken2011}
{Kuijken}, K. 2011, The Messenger, 146, 8

\bibitem[{{Lindegren} {et~al.}(2018){Lindegren}, {Hern{\'a}ndez}, {Bombrun},
  {Klioner}, {Bastian}, {Ramos-Lerate}, {de Torres}, {Steidelm{\"u}ller},
  {Stephenson}, {Hobbs}, {Lammers}, {Biermann}, {Geyer}, {Hilger}, {Michalik},
  {Stampa}, {McMillan}, {Casta{\~n}eda}, {Clotet}, {Comoretto}, {Davidson},
  {Fabricius}, {Gracia}, {Hambly}, {Hutton}, {Mora}, {Portell}, {van Leeuwen},
  {Abbas}, {Abreu}, {Altmann}, {Andrei}, {Anglada}, {Balaguer-N{\'u}{\~n}ez},
  {Barache}, {Becciani}, {Bertone}, {Bianchi}, {Bouquillon}, {Bourda},
  {Br{\"u}semeister}, {Bucciarelli}, {Busonero}, {Buzzi}, {Cancelliere},
  {Carlucci}, {Charlot}, {Cheek}, {Crosta}, {Crowley}, {de Bruijne}, {de
  Felice}, {Drimmel}, {Esquej}, {Fienga}, {Fraile}, {Gai}, {Garralda},
  {Gonz{\'a}lez-Vidal}, {Guerra}, {Hauser}, {Hofmann}, {Holl}, {Jordan},
  {Lattanzi}, {Lenhardt}, {Liao}, {Licata}, {Lister}, {L{\"o}ffler},
  {Marchant}, {Martin-Fleitas}, {Messineo}, {Mignard}, {Morbidelli}, {Poggio},
  {Riva}, {Rowell}, {Salguero}, {Sarasso}, {Sciacca}, {Siddiqui}, {Smart},
  {Spagna}, {Steele}, {Taris}, {Torra}, {van Elteren}, {van Reeven}, \&
  {Vecchiato}}]{Lindegren2018}
{Lindegren}, L., {et~al.} 2018, \aap, 616, A2

\bibitem[{{Liu} {et~al.}(2014){Liu}, {Yuan}, {Huo}, {Deng}, {Hou}, {Zhao},
  {Zhao}, {Shi}, {Luo}, {Xiang}, {Zhang}, {Huang}, \& {Zhang}}]{Liu2014}
{Liu}, X.-W., {et~al.} 2014, in IAU Symposium, Vol. 298, IAU Symposium, ed.
  S.~{Feltzing}, G.~{Zhao}, N.~A. {Walton}, \& P.~{Whitelock}, 310--321

\bibitem[{{Liu} {et~al.}(2019){Liu}, {Cui}, {Liu}, {Huang}, {Zhao}, \&
  {Zhang}}]{Liu2019}
{Liu}, Z., {Cui}, W., {Liu}, C., {Huang}, Y., {Zhao}, G., \& {Zhang}, B. 2019,
  \apjs, 241, 32

\bibitem[{{Ma{\'{\i}}z Apell{\'a}niz}(2005)}]{Maz2005}
{Ma{\'{\i}}z Apell{\'a}niz}, J. 2005, in ESA Special Publication, Vol. 576, The
  Three-Dimensional Universe with Gaia, ed. C.~{Turon}, K.~S. {O'Flaherty}, \&
  M.~A.~C. {Perryman}, 179

\bibitem[{{Ma{\'{\i}}z Apell{\'a}niz} {et~al.}(2008){Ma{\'{\i}}z
  Apell{\'a}niz}, {Alfaro}, \& {Sota}}]{Maz2008}
{Ma{\'{\i}}z Apell{\'a}niz}, J., {Alfaro}, E.~J., \& {Sota}, A. 2008, arXiv
  e-prints: 0804.2553

\bibitem[{{Ma{\'{\i}}z Apell{\'a}niz} \& {Barb{\'a}}(2018)}]{Maz2018b}
{Ma{\'{\i}}z Apell{\'a}niz}, J. \& {Barb{\'a}}, R.~H. 2018, \aap, 613, A9

\bibitem[{{Ma{\'{\i}}z Apell{\'a}niz} {et~al.}(2014){Ma{\'{\i}}z
  Apell{\'a}niz}, {Evans}, {Barb{\'a}}, {Gr{\"a}fener}, {Bestenlehner},
  {Crowther}, {Garc{\'{\i}}a}, {Herrero}, {Sana}, {Sim{\'o}n-D{\'{\i}}az},
  {Taylor}, {van Loon}, {Vink}, \& {Walborn}}]{Maz2014}
{Ma{\'{\i}}z Apell{\'a}niz}, J., {et~al.} 2014, \aap, 564, A63

\bibitem[{{Ma{\'{\i}}z Apell{\'a}niz} {et~al.}(2016){Ma{\'{\i}}z
  Apell{\'a}niz}, {Sota}, {Arias}, {Barb{\'a}}, {Walborn},
  {Sim{\'o}n-D{\'{\i}}az}, {Negueruela}, {Marco}, {Le{\~a}o}, {Herrero},
  {Gamen}, \& {Alfaro}}]{Maz2016}
{Ma{\'{\i}}z Apell{\'a}niz}, J., {et~al.} 2016, \apjs, 224, 4

\bibitem[{{Ma{\'{\i}}z Apell{\'a}niz} {et~al.}(2013){Ma{\'{\i}}z
  Apell{\'a}niz}, {Sota}, {Morrell}, {Barb{\'a}}, {Walborn}, {Alfaro}, {Gamen},
  {Arias}, \& {Gallego Calvente}}]{Maz2013}
{Ma{\'{\i}}z Apell{\'a}niz}, J., {et~al.} 2013, in Massive Stars: From alpha to
  Omega, 198

\bibitem[{{Ma{\'{\i}}z Apell{\'a}niz} {et~al.}(2011){Ma{\'{\i}}z
  Apell{\'a}niz}, {Sota}, {Walborn}, {Alfaro}, {Barb{\'a}}, {Morrell}, {Gamen},
  \& {Arias}}]{Maz2011}
{Ma{\'{\i}}z Apell{\'a}niz}, J., {Sota}, A., {Walborn}, N.~R., {Alfaro}, E.~J.,
  {Barb{\'a}}, R.~H., {Morrell}, N.~I., {Gamen}, R.~C., \& {Arias}, J.~I. 2011,
  in Highlights of Spanish Astrophysics VI, ed. M.~R. {Zapatero Osorio},
  J.~{Gorgas}, J.~{Ma{\'{\i}}z Apell{\'a}niz}, J.~R. {Pardo}, \& A.~{Gil de
  Paz}, 467--472

\bibitem[{{Ma{\'\i}z Apell{\'a}niz} \& {Weiler}(2018)}]{Maz2018}
{Ma{\'\i}z Apell{\'a}niz}, J. \& {Weiler}, M. 2018, \aap, 619, A180

\bibitem[{{Marigo} {et~al.}(2017){Marigo}, {Girardi}, {Bressan}, {Rosenfield},
  {Aringer}, {Chen}, {Dussin}, {Nanni}, {Pastorelli}, {Rodrigues}, {Trabucchi},
  {Bladh}, {Dalcanton}, {Groenewegen}, {Montalb{\'a}n}, \& {Wood}}]{Marigo2017}
{Marigo}, P., {et~al.} 2017, \apj, 835, 77

\bibitem[{{May} {et~al.}(1997){May}, {Alvarez}, \& {Bronfman}}]{May1997}
{May}, J., {Alvarez}, H., \& {Bronfman}, L. 1997, \aap, 327, 325

\bibitem[{{McClure-Griffiths} {et~al.}(2004){McClure-Griffiths}, {Dickey},
  {Gaensler}, \& {Green}}]{Mcclure2004}
{McClure-Griffiths}, N.~M., {Dickey}, J.~M., {Gaensler}, B.~M., \& {Green},
  A.~J. 2004, \apjl, 607, L127

\bibitem[{{Mohr-Smith} {et~al.}(2015){Mohr-Smith}, {Drew}, {Barentsen},
  {Wright}, {Napiwotzki}, {Corradi}, {Eisl{\"o}ffel}, {Groot}, {Kalari},
  {Parker}, {Raddi}, {Sale}, {Unruh}, {Vink}, \& {Wesson}}]{Mohr2015}
{Mohr-Smith}, M., {et~al.} 2015, \mnras, 450, 3855

\bibitem[{{Mohr-Smith} {et~al.}(2017){Mohr-Smith}, {Drew}, {Napiwotzki},
  {Sim{\'o}n-D{\'{\i}}az}, {Wright}, {Barentsen}, {Eisl{\"o}ffel}, {Farnhill},
  {Greimel}, {Mongui{\'o}}, {Kalari}, {Parker}, \& {Vink}}]{Mohr2017}
{Mohr-Smith}, M., {et~al.} 2017, \mnras, 465, 1807

\bibitem[{{Reed}(2003)}]{Reed2003}
{Reed}, B.~C. 2003, \aj, 125, 2531

\bibitem[{{Reid} {et~al.}(2014){Reid}, {Menten}, {Brunthaler}, {Zheng}, {Dame},
  {Xu}, {Wu}, {Zhang}, {Sanna}, {Sato}, {Hachisuka}, {Choi}, {Immer},
  {Moscadelli}, {Rygl}, \& {Bartkiewicz}}]{Reid2014}
{Reid}, M.~J., {et~al.} 2014, \apj, 783, 130

\bibitem[{{Russeil}(2003)}]{Russeil2003}
{Russeil}, D. 2003, \aap, 397, 133

\bibitem[{{Skiff}(2014)}]{Skiff2014}
{Skiff}, B.~A. 2014, VizieR Online Data Catalog, 1

\bibitem[{{Solomon} \& {Rivolo}(1989)}]{Solomon1989}
{Solomon}, P.~M. \& {Rivolo}, A.~R. 1989, \apj, 339, 919

\bibitem[{{Sun} {et~al.}(2017){Sun}, {Su}, {Zhang}, {Xu}, {Chen}, {Yang},
  {Jiang}, \& {Fang}}]{Sun2017}
{Sun}, Y., {Su}, Y., {Zhang}, S.-B., {Xu}, Y., {Chen}, X.-P., {Yang}, J.,
  {Jiang}, Z.-B., \& {Fang}, M. 2017, \apjs, 230, 17

\bibitem[{{Sun} {et~al.}(2015){Sun}, {Xu}, {Yang}, {Li}, {Du}, {Zhang}, \&
  {Zhou}}]{Sun2015}
{Sun}, Y., {Xu}, Y., {Yang}, J., {Li}, F.-C., {Du}, X.-Y., {Zhang}, S.-B., \&
  {Zhou}, X. 2015, \apjl, 798, L27

\bibitem[{{Vall{\'e}e}(2017)}]{Vallee2017}
{Vall{\'e}e}, J.~P. 2017, The Astronomical Review, 13, 113

\bibitem[{{Wall} \& {Jenkins}(2003)}]{Wall2003}
{Wall}, J.~V. \& {Jenkins}, C.~R. 2003, {Practical Statistics for Astronomers},
  ed. R.~{Ellis}, J.~{Huchra}, S.~{Kahn}, G.~{Rieke}, \& P.~B. {Stetson}

\bibitem[{{Weaver}(1970)}]{Weaver1970}
{Weaver}, H. 1970, in IAU Symposium, Vol.~38, The Spiral Structure of our
  Galaxy, ed. W.~{Becker} \& G.~I. {Kontopoulos}, 126

\bibitem[{{Xiang} {et~al.}(2017){Xiang}, {Liu}, {Yuan}, {Huo}, {Huang}, {Wang},
  {Chen}, {Ren}, {Zhang}, {Tian}, {Yang}, {Shi}, {Zhao}, {Li}, {Zhao}, {Cui},
  {Li}, {Hou}, {Zhang}, {Zhang}, {Wang}, {Wu}, {Cao}, {Yan}, {Yan}, {Luo},
  {Zhang}, {Bai}, {Yuan}, {Dong}, {Lei}, \& {Li}}]{Xiang2017}
{Xiang}, M.-S., {et~al.} 2017, \mnras

\bibitem[{{Xu} {et~al.}(2018{\natexlab{a}}){Xu}, {Bian}, {Reid}, {Li}, {Zhang},
  {Yan}, {Dame}, {Menten}, {He}, {Liao}, \& {Tang}}]{Xu2018}
{Xu}, Y., {et~al.} 2018{\natexlab{a}}, \aap, 616, L15

\bibitem[{{Xu} {et~al.}(2018{\natexlab{b}}){Xu}, {Hou}, \& {Wu}}]{Xu2018b}
{Xu}, Y., {Hou}, L.-G., \& {Wu}, Y.-W. 2018{\natexlab{b}}, Research in
  Astronomy and Astrophysics, 18, 146

\bibitem[{{Xu} {et~al.}(2006){Xu}, {Reid}, {Zheng}, \& {Menten}}]{Xu2006}
{Xu}, Y., {Reid}, M.~J., {Zheng}, X.~W., \& {Menten}, K.~M. 2006, Science, 311,
  54

\bibitem[{{Yuan} {et~al.}(2015){Yuan}, {Liu}, {Huo}, {Xiang}, {Huang}, {Chen},
  {Zhang}, {Sun}, {Wang}, {Zhang}, {Zhao}, {Luo}, {Shi}, {Li}, {Yuan}, {Dong},
  {Li}, {Hou}, \& {Zhang}}]{Yuan2015}
{Yuan}, H.-B., {et~al.} 2015, \mnras, 448, 855

\end{thebibliography}

\label{lastpage}
\end{document}